\documentclass[%
reprint,
superscriptaddress,
showpacs,preprintnumbers,
 amsmath,amssymb,
 aps,
prb,
floatfix,
]{revtex4-1}

\usepackage{graphicx}
\usepackage{dcolumn}
\usepackage{bm}

\usepackage{hyperref}
\hypersetup{
    unicode=false,          
    pdftoolbar=true,        
    pdfmenubar=true,        
    pdffitwindow=false,     
    pdfstartview={FitH},    
    pdftitle={My title},    
    pdfauthor={Author},     
    pdfsubject={Subject},   
    pdfcreator={Creator},   
    pdfproducer={Producer}, 
    pdfkeywords={keyword1} {key2} {key3}, 
    pdfnewwindow=true,      
    colorlinks=true,       
    linkcolor=blue,          
    citecolor=blue,        
    filecolor=blue,      
    urlcolor=blue,       
    plainpages=false        
}

\begin{document}
\title{A balancing act: Evidence for a strong subdominant $d$-wave pairing channel in ${\rm Ba_{0.6}K_{0.4}Fe_2As_2}$}
\date{\today}
\author{T. B\"ohm}
\affiliation{Walther Meissner Institute, Bavarian Academy of Sciences and Humanities, 85748 Garching, Germany}
\author{A.\,F. Kemper}
\affiliation{Lawrence Berkeley National Laboratory, 1 Cyclotron Road, Berkeley, CA 94720, USA}
\author{B. Moritz}
\affiliation{Stanford Institute for Materials and Energy Sciences, SLAC National Accelerator Laboratory,
2575 Sand Hill Road, Menlo Park, CA 94025, USA}
\author{F. Kretzschmar}
\affiliation{Walther Meissner Institute, Bavarian Academy of Sciences and Humanities, 85748 Garching, Germany}
\author{B. Muschler}
\affiliation{Walther Meissner Institute, Bavarian Academy of Sciences and Humanities, 85748 Garching, Germany}
\author{H.-M.~Eiter}
\affiliation{Walther Meissner Institute, Bavarian Academy of Sciences and Humanities, 85748 Garching, Germany}
\author{R. Hackl}
\affiliation{Walther Meissner Institute, Bavarian Academy of Sciences and Humanities, 85748 Garching, Germany}
\author{T.\,P. Devereaux}
\affiliation{Stanford Institute for Materials and Energy Sciences, SLAC National Accelerator Laboratory,
2575 Sand Hill Road, Menlo Park, CA 94025, USA}
\affiliation{Geballe Laboratory for Advanced Materials \& Dept. of Applied Physics,
Stanford University, CA 94305, USA}
\author{D.\,J. Scalapino}
\affiliation{Physics Department, University of California, Santa Barbara, CA 93106-9530, USA}
\author{Hai-Hu Wen}
\affiliation{National Laboratory of Solid State Microstructures and Department of Physics,
Nanjing University, Nanjing 210093, China}

\begin{abstract}
We present an analysis of the Raman spectra of optimally doped ${\rm Ba_{0.6}K_{0.4}Fe_2As_2}$ based on LDA band structure calculations and the subsequent estimation of effective Raman vertices. Experimentally a narrow, emergent mode appears in the $B_{1g}$ ($d_{x^2-y^2}$) Raman spectra only below $T_c$, well into the superconducting state and at an energy below twice the energy gap on the electron Fermi surface sheets. The Raman spectra can be reproduced quantitatively with estimates for the magnitude and momentum space structure of the s$_{+-}$ pairing gap on different Fermi surface sheets, as well as the identification of the emergent sharp feature as a Bardasis-Schrieffer exciton, formed as a Cooper pair bound state in a subdominant $d_{x^2-y^2}$ channel. The binding energy of the exciton relative to the gap edge shows that the coupling strength in this subdominant $d_{x^2-y^2}$ channel is as strong as 60\% of that in the dominant $s_{+-}$ channel. This result suggests that $d_{x^2-y^2}$ may be the dominant pairing symmetry in Fe-based sperconductors which lack central hole bands.
\end{abstract}
\maketitle


\section{Introduction}
Iron-based superconductors (FeSC) possess magnetically ordered spin-density wave (SDW) or possibly N\'eel order in close proximity to superconductivity \cite{Mazin:2008,Fernandes:2013,Liang:2013}. In general, the topology of the Fermi surface plays a crucial role in determining the type of order\cite{Hirschfeld:2011}. This sensitivity was demonstrated explicitly for single-layer FeSe\cite{HeSL:2013} which may become superconducting already above 60\,K possibly due to the interplay of intraband and interband Cooper pairing\cite{LeeJJ:2013}. In the superconducting phase, the structure, size, and potentially symmetry of the gap function $\Delta_{\bf k}$ is expected to react sensitively to small changes in external control parameters such as doping or pressure\cite{Thomale:2009,Thomale:2011}. Additionally, $\Delta_{\bf k}$ reflects the dominant channel for Cooper pairing and allows insight into unconventional pairing mechanisms driven by band structure dependent electronic interactions. Finding a way to monitor changes of the pairing state may provide a possible pathway for a quantitative description of superconductivity in the FeSCs. However, one needs the proper experimental tools.

One of the hallmarks of superconductivity in ${\rm Ba_{0.6}K_{0.4}Fe_2As_2}$ is the observation of  the neutron resonance\cite{Christianson:2008} which favors the $s_{+-}$ state predicted by Mazin and coworkers \cite{Mazin:2008}, but still leaves space for an $s_{++}$ state driven by orbital fluctuations \cite{Kontani:2010}. In either case, the nesting between the central hole bands and the electron bands takes advantage of strong interactions at short distances corresponding to a large momentum transfer at $(\pi,0)$ (in the 1\,Fe unit cell). Similarly, the electron bands themselves can gain from $(\pi,\pi)$ scattering of nearly equal strength\cite{Fernandes:2013,Liang:2013,Graser:2009,Graser:2010b}. Hence, two unconventional pairing states $s_{+-}$ and $d_{x^2-y^2}$ resulting from $(\pi,0)$ and $(\pi,\pi)$ scattering, respectively, can be expected to compete and may be tuned by intentionally changing the band structure.

The close proximity of these pairing instabilities leaves spectroscopic fingerprints. In the single particle spectra  one expects characteristic momentum dependence of the gaps on the Fermi surfaces of multi-band systems. If the gap changes sign between different sheets of the Fermi surface, the quasi-particle interference observed in tunneling spectra may demonstrate the influence from applied magnetic fields \cite{Hanaguri:2010}. In a light scattering experiment new or emergent collective modes are expected in addition to the more familiar pair breaking peak at an energy twice the gap maximum\cite{Devereaux:2007}. Generally these collective modes can appear in the particle-hole channel or particle-particle channel ($\tau_2$ or $\tau_3$ channels in the language of Nambu) either separately or together.

A critical question is whether these modes can be visible in Raman scattering measurements.
For example, narrow lines at lower energies originate from either residual excitonic interactions between the electrons of a broken pair\cite{Bardasis:1961,Klein:1984,Monien:1990,Chubukov:2009} or Josephson-like excitations \cite{Leggett:1966,Blumberg:2002,Klein:2010} between different bands in multi-band systems. In addition, there may be modes associated with coupled amplitude fluctuations of the superconducting and density wave gaps when charge density wave (CDW) ordering occurs\cite{Einzel:2013,Sooryakumar:1980,Littlewood:1981,Tutto:1992,Varma:2002,Barlas:2013}. A coupling between the superconducting and CDW channels allows a $\tau_2$ collective amplitude mode to be visible in Raman scattering measurements.

In both NbSe$_2$ and the A15 compounds V$_3$Si and Nb$_3$Sn, resolution-limited lines below twice the gap edge have been observed\cite{Sooryakumar:1980,Hackl:1983}, but the only evidence for amplitude modes was the approximate conservation of the integrated spectral weight of the in-gap mode and the phonon-like excitation either as a function of applied field \cite{Sooryakumar:1980} or temperature \cite{Littlewood:1981,Hackl:1983,Hackl:1989,Measson:2014}. There are no systematic studies on Leggett modes but the data in MgB$_2$ suggest that there is a mode in the right range of energy which originates from a weak coupling between the two-dimensional (2D) $\sigma$ band which possesses a large gap and the more 3D $\pi$ band \cite{Blumberg:2002,Klein:2010}.

First predicted by Bardasis and Schrieffer (BS) \cite{Bardasis:1961,Klein:1984,Monien:1990}, depending on the sign of the residual interaction, excitonic or electron pair bound states can be formed. These excitonic or electron pair modes may exist as sharp features below the gap in $s-$wave superconductors or in $d-$wave superconductors they may be
damped considerably due to the existence of quasiparticles from the presence of gap nodes\cite{Devereaux:1995a}.
BS modes have been observed in superfluid $^4$He \cite{Greytak:1969} where they correspond to bound pairs of rotons\cite{Zawadowski:1972}, and could be an alternative explanation for the in-gap modes in A15 compounds V$_3$Si and Nb$_3$Sn \cite{Monien:1990}. In both of these compounds, structural transitions from a high temperature-cubic to a low-temperature-tetragonal
lattice occur above the superconducting transition temperature \cite{Rehwald:1972}, but no evidence of a CDW appears at lower temperatures. Nevertheless a spectral weight transfer from the phonon into the collective modes appears below the superconducting transition temperature, similar to the case of NbSe$_2$.

Recently, narrow in-gap modes were observed in ${\rm Ba_{0.6}K_{0.4}Fe_2As_2}$ and interpreted in terms of BS modes \cite{Kretzschmar:2013}. In this case, the intensity does not come from a phonon, some of which gain rather than lose spectral weight upon entering the superconducting state, but is drained from the pair-breaking peaks. This experimental observation is qualitatively different from what was found in previous studies \cite{Devereaux:2007}, but the effect was predicted specifically for the iron-based compounds with competing $s-$ and $d-$wave pairing states \cite{Scalapino:2009}. Therefore, the earlier qualitative argumentation \cite{Kretzschmar:2013} needs to be augmented both experimentally and theoretically.

In this work we present experimental polarization-dependent Raman spectra for various temperatures between the low-temperature limit of approximately 8\,K and 46\,K. In addition, we performed weak coupling calculations for $T=0$ on a realistic band structure taking into account interactions between the five bands close to the Fermi level. These theoretical results and observation of a BCS-like temperature dependence of an emergent mode at 140\,cm$^{-1}$ allow us to uniquely identify it as a $d_{x^2-y^2}$ BS exciton. From the energy position and the spectral weight of the exciton, we estimate the relative strength of the subdominant $d_{x^2-y^2}$ pairing channel to be more than half as strong as the dominant $s_{+-}$ channel. Our results suggest that the $d_{x^2-y^2}$ pairing channel may indeed become dominant when the $s_{+-}$ interaction is reduced, for example, by the absence of hole pockets at the center of the Brillouin zone.

\section{Experimental study of the temperature dependence}
In ${\rm Ba_{0.6}K_{0.4}Fe_2As_2}$, superconductivity-induced features were found in all symmetries experimentally accessible with light polarizations in the Fe planes\cite{Kretzschmar:2013}. Although the crystal unit cell involves 2 iron atoms per unit cell due to the staggered positioning of the arsenic above and below the Fe planes, it is convenient to instead make group theory reference in the 1 Fe unit cell where polarizations and selection rules can be framed in terms of polarizations along the Fe-Fe bond direction. In the $B_{1g}$ spectra (1\,Fe unit cell) the observation of very narrow modes at low temperature suggests the existence of collective excitations and their interpretation in terms of excitonic BS modes. However, the temperature dependence or the energy and emergence of this mode requires further study.

Before describing the quantitative theoretical analysis in the zero-temperature limit, we present an additional set of experiments in the range $0<T\le 46$\,K, since we expect that the in-gap modes and the usual pair-breaking features depend differently on temperature in systems with intermediate to strong coupling. While the in-gap modes should by and large follow the temperature dependence of the single-particle gap\cite{Monien:1990}, interactions that give rise to Raman peaks in the normal state reduce the temperature dependence of the pair breaking features\cite{Devereaux:1993,Manske:2004}.

The experiments were performed on a freshly cleaved surface of the same optimally hole-doped single crystal of $\mathrm{Ba_{0.6}K_{0.4}Fe_2As_2}$ which has been used in previous studies and possesses a superconducting transition at $T_c=38.5$\,K\cite{Kretzschmar:2013}. We used an Ar ion laser emitting at 514\,nm and a standard scanning spectrometer with the sample held in a cryogenically pumped vacuum. We measured spectra with linear polarizations of the incoming and outgoing photons oriented perpendicular and at 45$^{\circ}$ with respect to the Fe--Fe direction (0$^{\circ}$ w.r.t. the crystallographic axes) to project the $B_{1g}$ and $A_{2g}$ symmetries. No subtraction procedure was applied as the $A_{2g}$ component was found to be weak. The $B_{1g}$ spectra contain all relevant features.

\begin{figure}[tbp]
  \centering
  \includegraphics[width=1.0\columnwidth]{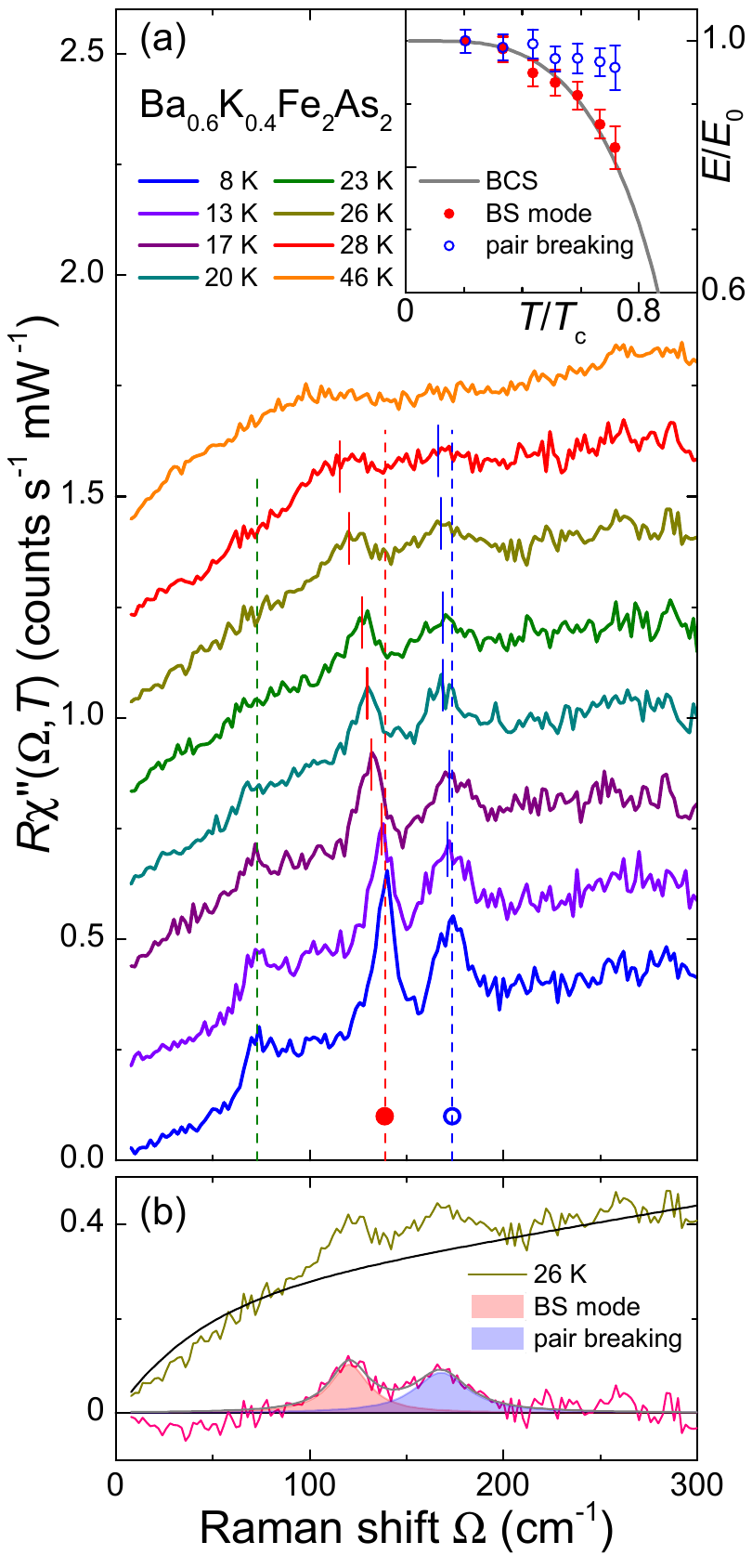}
  \caption{Temperature dependence of the Raman spectra of $\mathrm{Ba_{0.6}K_{0.4}Fe_2As_2}$ in $B_{1g}$ symmetry. (a) The spectra measured above 8\,K are consecutively shifted up by 0.2 units. The pair-breaking features (open symbols) and the collective mode (full circles) depend differently on temperature as shown in the inset (having a suppressed energy zero). The pair-breaking maximum exhibits a temperature dependence that is different from BCS due to interaction effects\cite{Devereaux:1993,Manske:2004}. (b) The peak energies were determined by fitting the spectra with two Lorentzians and a smooth background.
  }
  \label{fig:T}
\end{figure}

The spectra measured at various temperatures between 8\,K and 46\,K are shown in Fig.~\ref{fig:T}\,(a). At low temperature one observes two prominent peaks at 140 and 170\,cm$^{-1}$ and a weak one at 70\,cm$^{-1}$ which were previously identified with collective modes \cite{Kretzschmar:2013}. The mode at 140\,cm$^{-1}$ has the smallest low-temperature width. Following the positions of the three peaks (dashed vertical lines in Fig.~\ref{fig:T}\,(a) show the low-temperature limit) indicates distinct differences, with the mode at 140\,cm$^{-1}$ displaying the strongest shift and the peaks at 70\,cm$^{-1}$  and the gap edge at 170\,cm$^{-1}$ (open circle) varying only weakly. The positions are determined following a background subtraction (Fig.~\ref{fig:T}\,(b)). In the inset of Fig.~\ref{fig:T}\,(a), the positions of the two high-energy peaks relative to their low-temperature limiting values are shown along with the BCS prediction for the energy gap. Only the mode at 140\,cm$^{-1}$ is close to the mean-field expectation in striking similarity with the single-particle gaps\cite{Evtushinsky:2009}, and we conclude that the mode at 140\,cm$^{-1}$ is the only candidate for a BS exciton. With increasing temperature the width of the line increases due to quasiparticle damping making it indiscernible at sample temperatures above 28\,K.

The additional experimental observation of the temperature dependence facilitates a clear distinction between the various spectral features and motivates us to explain the 70 and 170\,cm$^{-1}$ modes in terms of pair-breaking and identify only the line at 140\,cm$^{-1}$ with a BS exciton. This considerably simplifies the calculations. Nevertheless, it remains crucially important to work with a realistic band structure, since the vertex corrections result from interband terms \cite{Scalapino:2009} rather than from intraband anisotropies of the interaction potential $V_{{\bf k},{\bf k}^{\prime}}$ as derived first by Bardasis and Schrieffer \cite{Bardasis:1961} and discussed in detail later in the context of light scattering \cite{Klein:1984,Monien:1990}.

\section{Model Description}
The model employed in this study is based on a realistic tight-binding bandstructure derived from DFT/LDA estimates \cite{Graser:2010b} to provide a quantitative analysis of the Raman spectra for optimally hole doped $\mathrm{Ba_{0.6}K_{0.4}Fe_2As_2}$ in the superconducting state. Since the experiments show evidence of a bound state inside the gap in addition to the superconducting pair-breaking features, vertex corrections must be considered. Physically the vertex corrections describe the final-state interaction between the two electrons of a Cooper pair which have been broken by a photon such that the model accounts for both the pair-breaking effect and the final-state interaction on a realistic 3D multi-band tight-binding bandstructure.

The band structure is generated from a tight-binding approximation to the five Fe $d$-orbitals developed by Graser \textit{et al.} \cite{Graser:2010b} for undoped $\mathrm{BaFe_2As_2}$, with the Fermi energy shifted down by 144\,meV with respect to the original bandstructure to account for the substitution of 40\% Ba by K which adds 0.2 holes per Fe atom and reduces the filling to 5.8. Transforming the system from an orbital basis to a band basis gives five bands of which four cross the Fermi level including the two hole bands in the Brillouin zone (BZ) center, one hole band at the $M$-point, and an electron band encircling the $X$-point. The presence of the 2\,Fe unit cell requires a backfolding of the 1\,Fe BZ, achieved by adding another five bands, shifted by ${\bf k} = (\pi,\pi,\pi)$, to the existing ones. This vector accounts for the additional translational symmetry of the 2\,Fe BZ. Five of the resulting ten bands cross the Fermi level: three hole bands in the BZ center (h1, h2, and h3 from the inside out) and two electron bands around the $X$-point (e1 and e2 from the outside in). The Fermi surfaces of the hole bands intersect each other on lines as do the electron bands. Since the intersecting bands derive from the same orbitals the degeneracies are lifted\cite{Mazin:2010a} by any small residual interaction: we used 25\,meV for all bands and show later that the hybridization energy influences the Raman spectra only weakly.

\begin{figure}[h!]
  \centering
  \includegraphics[width=0.75\columnwidth]{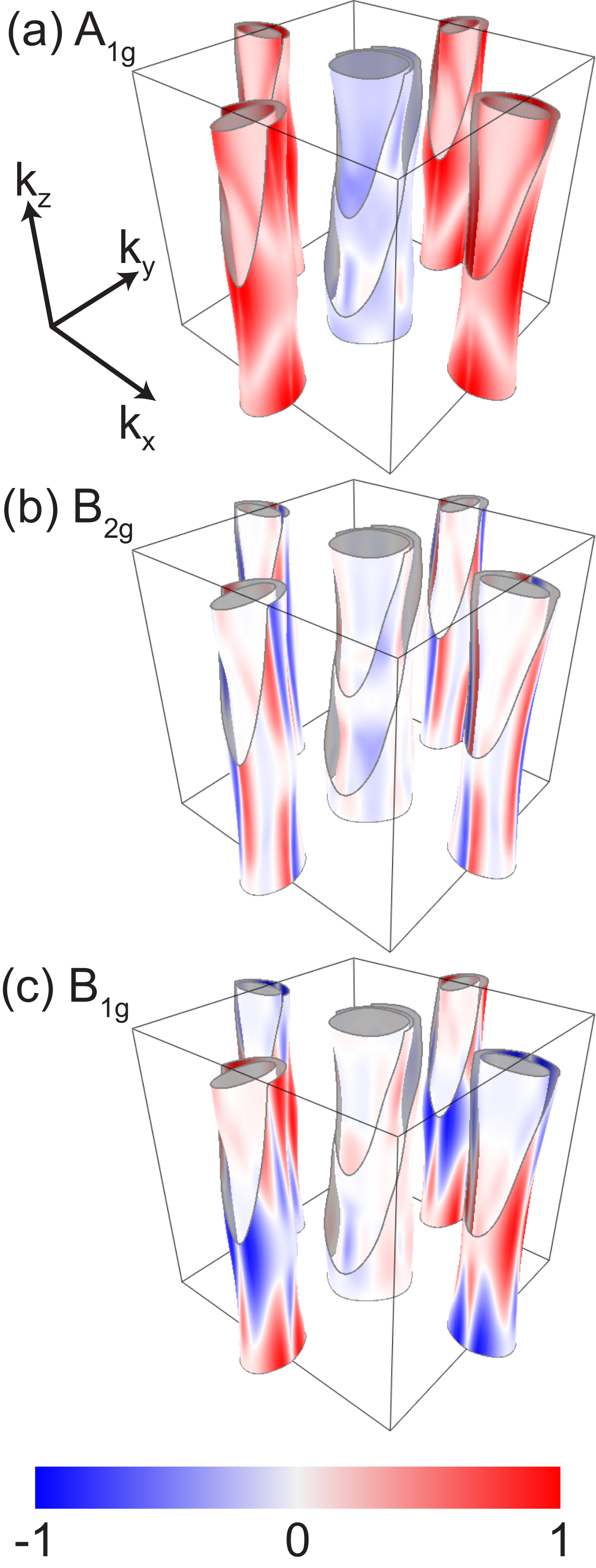}
  \caption{Raman vertices in (a) $A_{1g}$, (b) $B_{2g}$, and (c) $B_{1g}$ symmetry. The black frame represents the 1\,Fe BZ  ranging from $-\pi$ to $\pi$ in each dimension $k_x$, $k_y$, and $k_z$. There are three Fermi surfaces in the center and two at each face (the outer ones are cut open to visualize the inner ones) showing the hole bands and electron bands, respectively. All three symmetries have a common color scale that shows the sign and intensity of the Raman vertex at the Fermi surface. The hole bands around the BZ edges are equivalent to those in the center and are omitted for simplicity.
  }
  \label{fig:vertices}
\end{figure}

For calculating the Raman response the momentum dependent vertices are needed \cite{Devereaux:1994,Devereaux:2007}. This is tractable only in the effective-mass approximation, as justified here \cite{Mazin:2010a}, since the incident photons are lower than resonance energies. The related vertices $\gamma_{n}^{\mu}$  for symmetry $\mu$ ($A_{1g}$, $B_{1g}$, $B_{2g}$)  are derived numerically from the dispersion $E_{n}(\mathbf{k})$ of band $n$, given by
\begin{equation}
  \gamma_{n}^{A_{1g}}(\mathbf{k}) =
  \frac{1}{2} \left\{ \frac{\partial^2 E_{n}(\mathbf{k})}{\partial k_{x} \partial k_{x}} +
  \frac{\partial^2 E_{n}(\mathbf{k})}{\partial k_{y} \partial k_{y}} \right\},
\label{eq:ramanverticesA1g}
\end{equation}

\begin{equation}
  \gamma_{n}^{B_{1g}}(\mathbf{k}) =
  \frac{1}{2} \left\{ \frac{\partial^2 E_{n}(\mathbf{k})}{\partial k_{x} \partial k_{x}} -
  \frac{\partial^2 E_{n}(\mathbf{k})}{\partial k_{y} \partial k_{y}} \right\},
\label{eq:ramanverticesB1g}
\end{equation}
\begin{equation}
  \gamma_{n}^{B_{2g}}(\mathbf{k})=\frac{\partial^2 E_{n}(\mathbf{k})}{\partial k_{x} \partial k_{y}},
\label{eq:ramanverticesB2g}
\end{equation}
and shown in Fig.~\ref{fig:vertices}. Here $k_{x,y}$ refer to momenta along Fe-Fe bond directions.
Although the bands are calculated for the 2\,Fe BZ, we continue to use the 1\,Fe BZ reference frame because the experiments clearly show that the symmetry selection rules are dominated by the 1\,Fe cell \cite{Muschler:2009,Chauviere:2010,Kretzschmar:2013} and because in the 2\,Fe BZ cell the role of the $B_{1g}$ and $B_{2g}$ projections would be interchanged in a counterintuitive way such that the $d_{x^2-y^2}$ type of interaction between the electron bands would appear in the $B_{2g}$ channel. In fact, the 1\,Fe BZ captures most of the features and simplifies the argumentation considerably while for the 2\,Fe BZ improvements  are found only on a quantitative level while the results are qualitatively similar.

Figures~\ref{fig:vertices}\,(b) and (c) show directly that the strongest contributions for the $B_{1g}$ and $B_{2g}$ spectra come from the outer electron band. For a more quantitative statement, the contribution from superconductivity to the Raman response $\chi^{\prime\prime}(\mathbf{q}=0,\Omega) =\mathrm{Im}\chi(\mathbf{q}=0,\Omega)$ is evaluated by an intraband bare bubble approximation,
\begin{equation}
\chi(\Omega) = \sum_{n} \sum_{\mathbf{k}} \gamma^2_n(\mathbf{k}) \lambda_n(\mathbf{k},\Omega)
\label{eq:responsescksum}
\end{equation}
where the  $\lambda_n(\mathbf{k},\Omega)$ is given by the Tsuneto function\cite{Tsuneto:1960}, where $\Omega$ is the Raman shift. Neglecting band structure effects the expression for the response at $T=0$ can be transformed into
\begin{equation}
\chi^{\prime\prime}(\Omega) = 4 \pi \sum_{n}\langle \frac{\gamma_{n}^2(\mathbf{k}) |2\Delta_{n}(\mathbf{k})|^2}{\Omega\sqrt{\Omega^2 - |2\Delta_{n}(\mathbf{k})|^2  }}\rangle
\label{eq:responsesc}
\end{equation}
where $\langle\dots\rangle$ denotes an average over Fermi surface sheet $n$. The only relevant physical parameters which are varied to achieve the best agreement with the data are $\mathbf{k}$-dependent gap structures $\Delta_n(\mathbf{k})$ for each Fermi surface. In addition, the relative intensities of the spectra are scaled by 0.3, 0.6, and 1 for $A_{1g}$, $B_{1g}$, and $B_{2g}$, respectively. For the $A_{1g}$ spectra screening is included \cite{Devereaux:1996}, but the effects are found to be very small since the gaps on the electron and hole bands are quite symmetric (except for the outer hole band), and the concomitant sign change of the Raman vertex nearly cancels all the screening contributions\cite{Boyd:2009,Scalapino:2009}.

In addition to the response at lowest order, corrections from the final state interaction between the two single electrons created by Cooper pair breaking by photons have to be considered \cite{Klein:1984,Monien:1990}. The dynamics of bound states becomes important whenever there are anisotropies in the pairing potential \cite{Bardasis:1961} corresponding to interactions beyond ground state Cooper pairing. Since this competition is important in the FeSCs because of bands at high-symmetry points, we evaluated higher orders of perturbation theory (vertex corrections).

Here, contributions  originating from a $d$-wave attractive coupling between the outer electron bands will be included that lead to a collective excitonic mode in $B_{1g}$ symmetry\cite{Scalapino:2009}. The additional coupling $g(\mathbf{k})$ contributes to the anisotropy of $V_{{\bf k},{\bf k}^{\prime}}$ and is assumed to be relevant only between the outer electron bands (e1 in Table~\ref{tab:ARPES}). The influence on the response from e2 has been found to be negligible. We further assume that $V_{{\bf k},{\bf k}^{\prime}}$ is separable and varies as $g(\mathbf{k})\lambda_d g(\mathbf{k^\prime})$ with $g(\mathbf{k})$ proportional to $\gamma^{B_{1g}}(\mathbf{k})$ and normalized in a way that $\lambda_d$ measures the strength of the $d$-wave interaction. $g(\mathbf{k})$ causes multiple scattering processes and leads to an additive term in the response of the outer electron bands as discussed in detail in Ref. \onlinecite{Scalapino:2009} where
\begin{equation}
  \Delta\chi^{\prime\prime}(\Omega) = \left(\frac{2}{\Omega}\right)^2 \mathrm{Im} \left\{\frac{\langle \gamma(\mathbf{k}) g(\mathbf{k})  \Delta(\mathbf{k})\bar{P}(\Omega, \mathbf{k})  \rangle^2}
  {\left( \lambda_d^{-1} - \lambda_s^{-1} \right) - \langle g^2 \bar{P}(\Omega, \mathbf{k}) \rangle} \right\}.
\label{eq:responsecm}
\end{equation}
$\bar{P}(\Omega, \mathbf{k})$ is the response kernel (see Eq.~(12) of Ref.\,\onlinecite{Scalapino:2009} or, for isotropic systems, Eqs.~(B6a)-(B6c) in Ref.\,\onlinecite{Monien:1990}) and $\lambda_s$ is the average coupling in the dominant $s$-wave ground state. $\lambda_d$ is expressed as a fraction of $\lambda_s$. Note that the vertex $\gamma(\mathbf{k})$ appears only linearly inside the Fermi surface average $\langle\dots\rangle$.

\section{Results and Discussion}
\begin{figure}[tbp]
  \centering
  \includegraphics[width=1.0\columnwidth]{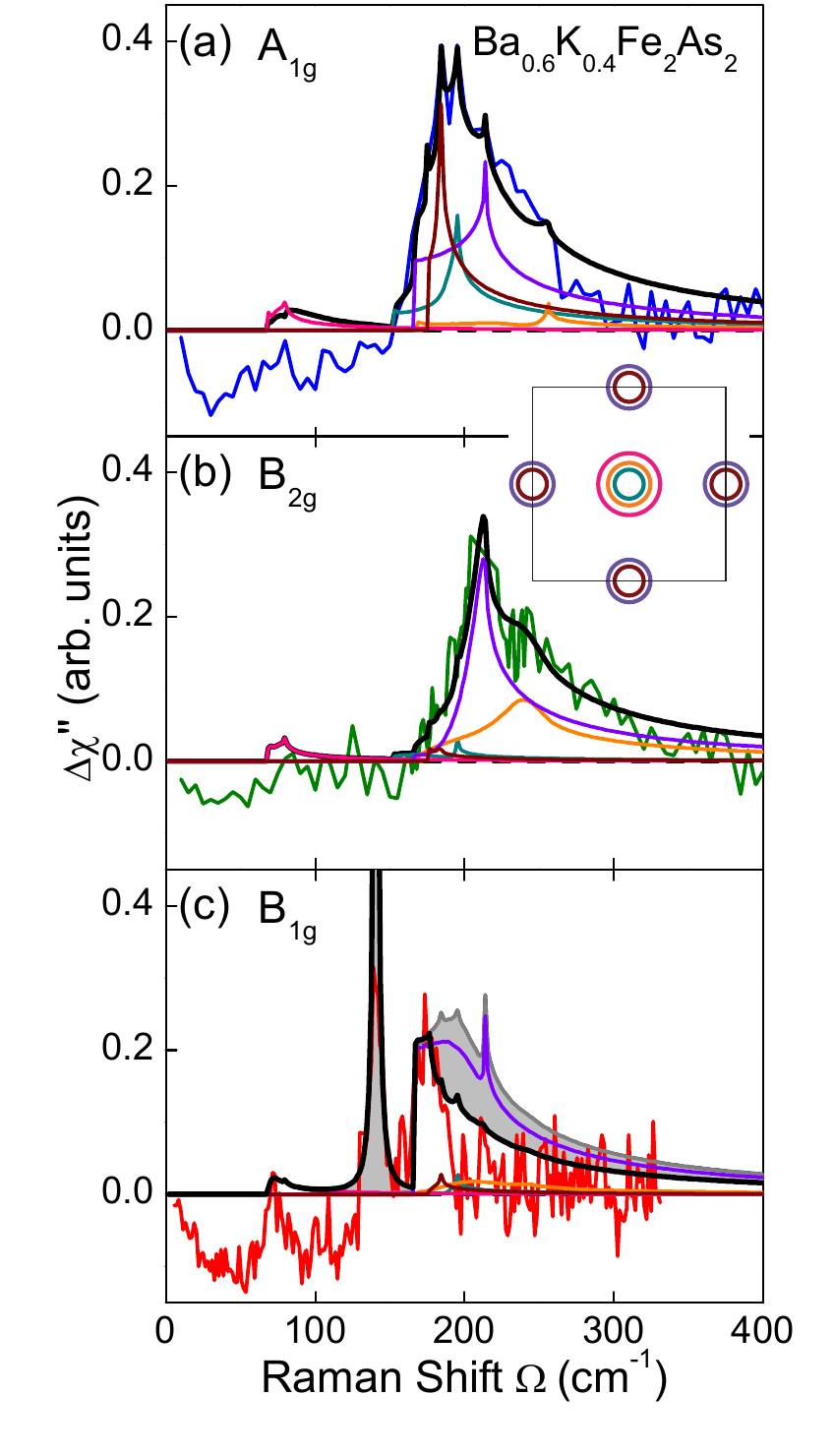}
  \caption{Raman response and theoretical results for (a) $A_{1g}$, (b) $B_{2g}$, and (c) $B_{1g}$ symmetry in $\mathrm{Ba_{0.6}K_{0.4}Fe_2As_2}$. Shown is the difference $\Delta\chi^{\prime\prime}$ of the response at 8 and 45\,K (raw data see Ref.\,\onlinecite{Kretzschmar:2013}). In this way temperature independent phonon lines and the particle-hole continuum are subtracted out not being described by the weak-coupling theory. As a side effect, the experimental intensities become negative inside the gap while the theoretical spectra just vanish. The inset between panels (a) and (b) shows a cartoon of the bands in the 1\,Fe zone for facilitating the identification of the response from each band via the color. The sum of the contributions is shown in black. (c) The grey shaded area is the spectral weight transferred from the pair-breaking region into the collective mode.}
	\label{fig:Fits}
\end{figure}

The model has been applied to the experimental data with the fitted results shown in Fig. \ref{fig:Fits} compared to experiments. The raw data \cite{Kretzschmar:2013} are a superposition of the electronic continuum and phonons. If the normal state spectra are subtracted from those in the superconducting state only superconductivity-induced features survive. If the phonons are not sensitive to the superconducting transition they disappear completely since the normal state temperature dependence is already too weak to be visible below 50\,K. In the case of the FeSCs most of the phonons are indeed weakly coupled \cite{Boeri:2008,Rahlenbeck:2009} and disappear here. Only the $B_{1g}$ Fe mode becomes more intense. The continuum at energies above twice the gap maximum consists only of superconductivity-induced changes thus simplifying the comparison with weak-coupling results. In the gap region the difference spectra become negative, but the theoretical predictions yield vanishing intensity with negligible conceptual complications.

All three symmetries show depletion of spectral weight in the low energy region and an enhancement for energies larger than about 160 cm$^\mathrm{-1}$. This behavior clearly indicates the existence of a superconducting gap. Another common feature is a small enhancement at about 70 cm$^\mathrm{-1}$, however, the sharp peak at 140 cm$^\mathrm{-1}$ in the $B_{1g}$ channel, which is almost resolution limited, does not have a correspondence in the other symmetries and can be identified as a $d$-wave collective mode. The following quantitative analysis is designed to support this interpretation and to reveal properties of both the superconducting gaps and the collective mode.

An ideal starting point for analysis is the $B_{2g}$ spectrum which is free of collective modes and screening effects. Eq.~\eqref{eq:responsesc} can be applied separately for each band with additive results. One finds that only the contributions from the outer electron band (e1, purple in Fig.~\ref{fig:Fits}) and the middle hole band (h2, orange) are large enough to contribute significantly to the response above 160\,cm$^\mathrm{-1}$, with the contribution from the outer electron band approximately twice as large as that of the middle hole band. This difference can be anticipated just by looking at the $B_{2g}$ Raman vertices of Fig.~\ref{fig:vertices} with a high intensity on the outer electron band, a smaller intensity for the middle hole band and vanishingly small intensities from the other bands. To fully reproduce the increase of the spectrum between 160\,cm$^\mathrm{-1}$ and the maximum at 210\,cm$^\mathrm{-1}$  it is necessary to (i) adjust the minimum and maximum gap values on band e1 and (ii) align the gap minimum with the minimum of the Raman vertex. This alignment allows one to reproduce the experimental slope without a spectral discontinuity. For band e1, one assumes that the gap has four-fold symmetry, with the maxima aligned along the $k_{x}$ and $k_{y}$ directions, and with no $k_{z}$ dispersion for the fit. The remaining shoulder on the high-energy side of the peak can be reproduced with a $k_{z}$ dispersive gap on the middle hole band having the maximum and the minimum at $k_{z}=0$ and $k_{z}=\pm \pi$, respectively. The gap maxima and minima and the functional variations along $k_z$ and in the basal plane are given in Table~\ref{tab:ARPES}. The black line in Fig.~\ref{fig:Fits}\,(b) is the sum of all contributions.

\begin{figure}[htbp]
  \centering
  \includegraphics[width=1.0\columnwidth]{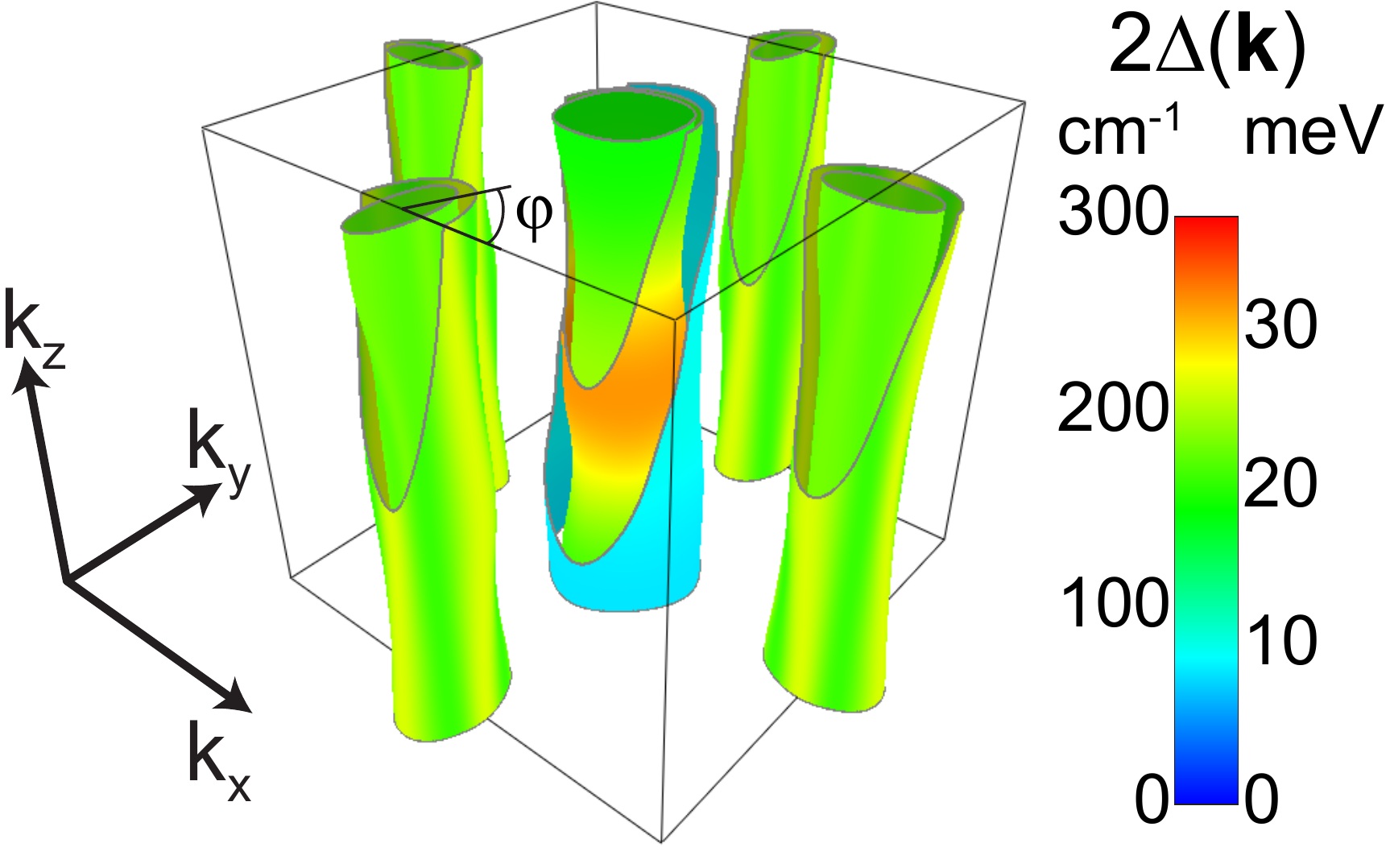}
  \caption{Magnitudes of the band-dependent gaps $2\Delta_n(\bf k)$ at the Fermi momentum $\bf k_F$ shown in false color. A moderate gap anisotropy is found on the middle hole band h2 with the absolute gap maximum at $k_{z}=0$.  The gaps on the electron bands vary with the azimuthal angle $\varphi$ and have maxima along $k_{x}$ and $k_{y}$. The gap anisotropies on the inner (h1) and outer (h3) hole bands and on the inner electron band (e2) are negligible.
  }
  \label{fig:Gaps}
\end{figure}

For the $A_{1g}$ spectrum the inner hole band (h1, dark cyan) and the inner electron band (e2, brown) become important (Fig. 1a). However, neither band can be expected to produce a feature at 70\,cm$^\mathrm{-1}$ because the contributions from h1 and e2 would be too large in $A_{1g}$ symmetry, but too small in the $B_{1g}$ spectrum. Hence, the outer hole band (h3, pink), for which the nesting condition is worse than for the other bands, is used to reproduce the feature at 70\,cm$^\mathrm{-1}$. The two remaining bands h2 and e1 are used to reproduce the shape in the 190\,cm$^\mathrm{-1}$ range being approximately 20\,cm$^\mathrm{-1}$ below the maximum in $B_{2g}$ symmetry. All gap magnitudes used for describing the experimental spectra (Fig.~\ref{fig:Fits}) are compiled in Table~\ref{tab:ARPES} and shown in false-color representation in Fig.~\ref{fig:Gaps} in the 1\,Fe reference frame. The gap is as large as $\Delta_{\rm h2}=15.9$\,meV on the middle hole band (h2) at $k_z=0$. The minimal gap is found on the outer hole band (h3). The gaps on the electron bands vary only in the $k_x$--$k_y$ plane.

\begin{table}[htbp]
  \caption{Gap energies (meV) as obtained by Raman scattering and angle resolved photoemission spectroscopy (ARPES) \cite{Nakayama:2009}. The bands h1--e2 are color coded in Fig.~\ref{fig:Fits} as follows: h1 (inner hole band) dark cyan, h2 (middle hole band) orange, h3 (outer hole band) pink, e1 (outer electron band) purple, e2 (inner electron band) brown. Bands h1 and h2 cannot be distinguished in the ARPES experiment and have therefore the same entry. In the analysis of the Raman data the gaps on the hole and the electron bands depend on $k_z$ ($-\pi<k_z<\pi$) and, respectively, on the azimuthal angle $\varphi$ as defined in Fig.~\ref{fig:Gaps} ($0<\varphi<2\pi$). $\bar\Delta_n$ and $\eta_n$ represent the average and the modulation amplitude of the gap on the Fermi surface of band $n$, respectively.}
  \vspace{1mm}
  \centering	
  \begin{tabular}{l r@{.}l r@{.}l c r@{.}l}
  \hline\hline\\[-2ex]
  Band & \multicolumn{2}{c}{$\Delta\mathrm{_{min}^{Raman}}$} & \multicolumn{2}{c}{$\Delta\mathrm{_{max}^{Raman}}$} & $\Delta({\bf k})$ & \multicolumn{2}{c}{$\Delta\mathrm{_{ARPES}}$}\\
  \hline\\[-2ex]
  h1 &  ~9&5 & ~12&1 &                                                               & 12&$3 \pm 0.6$\\
  h2 & ~10&4 & ~15&9 &                          $\bar\Delta_n+\eta_n\cos(k_z)$       & 12&$3 \pm 0.6$\\
  h3 &  ~4&2 &  ~5&0 &                                                               &  5&$8 \pm 0.8$\\
  e1 & ~10&3 & ~13&3 &                                                               & 12&$2 \pm 0.3$ \\
  e2 & ~10&8 & ~11&4 & \raisebox{1.5ex}[-1.5ex]{$\bar\Delta_n+\eta_n\cos(4\varphi)$} & 11&$4 \pm 0.5$ \\
  \hline
  \hline
  \end{tabular}	
  \label{tab:ARPES}
\end{table}
As opposed to the $A_{1g}$ and the $B_{2g}$ spectra, the $B_{1g}$ spectrum cannot be reproduced with the choice of gaps summarized in Table~\ref{tab:ARPES} and Fig.~\ref{fig:Gaps}. According to the Raman vertices only the outer electron band e1 contributes significantly while the intensity should be comparable both in $B_{2g}$ and $B_{1g}$ symmetries. To resolve this discrepancy the effect of an excitonic collective mode is introduced.

The subdominant coupling $g$ shifts spectral weight from the pair breaking peak into the sharp collective mode.  A momentum dependent $g(\mathbf{k})$ must be utilized which reduces the  response only at the gap maximum (rather than the minimum) while leading to the excitonic peak at 140\,cm$^\mathrm{-1}$. The best choice is a $d$-wave form for $g(\mathbf{k})$ which is small along diagonal directions to maximize the coupling between the gap maxima of the outer electron bands and, in addition, is proportional to the $B_{1g}$ vertex (Eq.~\ref{eq:ramanverticesB1g}). The latter specialization is necessary since the linear $B_{1g}$ vertex in Eq.~\eqref{eq:responsecm} has several sign changes [see Fig.~\ref{fig:vertices}\,(c)] and would nearly cancel the weight of the collective mode for a weakly $k$-dependent $g$. The reason for this artifact originates in the fine structure of the $B_{1g}$ vertex which enters to lowest order quadratically, but linearly in the vertex. The choice of $g(\mathbf{k}) \propto \gamma^{B_{1g}}(\mathbf{k})$ is physically justified and ensures that the spectral weights of both the bare bubble and the vertex correction come from the same parts of the Fermi surface. This argument is particularly relevant for comparing the two coupling channels.

In fact, the transfer of spectral weight encodes the relative strength between the $s$ and $d$ channels. In addition to the weight transfer, the position of the collective mode depends on $(\lambda_d/\lambda_s)^2$ as derived for an isotropic gap by Monien and Zawadowski\cite{Monien:1990}. With the maximal gap $2\Delta_{\rm max}$ of 210 cm$^{-1}$ on the e1 band and the collective mode at  140\,cm$^{-1}$, the binding energy is as large as one third of $2\Delta_{\rm max}$ yielding $\lambda_d \approx 0.6\lambda_s$. A similar ratio was used in the model calculations of Ref.\onlinecite{Scalapino:2009}. The related transfer of approximately one half of the spectral weight from the pair-breaking maximum into the bound state [Fig.~\ref{fig:Fits}] is consistent with the energy shift. This rather high fraction highlights that the subdominant $d$-wave channel lies in close proximity to the $s$-wave channel such that if the $s$-wave channel weakens, for instance as a result of a change of the Fermi surface, a new dominant symmetry emerges and a BS mode would flip identity as a subdominant $s-$wave bound state exciton. We speculate, that this could be realized in FeSe [\onlinecite{Hanaguri:2010}] and alkali-doped selenides such as ${\rm Rb_{0.8}Fe_{1.6}Se_2}$ although the pairing symmetry in these systems is still a matter of intense discussion.\cite{Qian:2011,Mazin:2011,Wang:2012,Khodas:2012,Kretzschmar:2013}

\section{Conclusions}
We studied the temperature dependence of the $B_{1g}$ Raman spectra in ${\rm Ba_{0.6}K_{0.4}Fe_2As_2}$ and proposed a realistic model calculation for the superconducting response at low temperatures that reproduces the spectra almost quantitatively.
The temperature dependence observed for the prominent peaks and the theoretical analysis demonstrate that only the $B_{1g}$ mode at 140\,cm$^{-1}$ has all features expected for a BS mode in the presence of competing pairing symmetries: it lies below twice the gap edge, has an almost resolution limited width, drains energy from the pair-breaking peaks, and has a temperature dependence which is dominated by that of the single-particle gap. The coupling parameter in the subdominant $d_{x^2-y^2}$ channel reaches 60\% of the prevailing $s$ pairing state making $d_{x^2-y^2}$ pairing a candidate for materials without central hole bands.

The $d_{x^2-y^2}$ channel competes with the $s$ ground state (independent of whether it is $s_{+-}$ or $s_{++}$) since the gaps on electron bands have the same sign whereas the $d_{x^2-y^2}$ channel would lead to a phase difference of $\pi$ between neighboring electron bands. Although the $d$ channel is already quite strong in ${\rm Ba_{0.6}K_{0.4}Fe_2As_2}$ the gaps on the various Fermi surfaces are not very anisotropic yet. For this reason $T_c$ is relatively high, and the density of states between the large and the small gaps is sufficiently small on the relevant bands thus keeping the damping of the excitonic mode small. If the ratio of the coupling strengths comes closer to one the frustration between the $s$ and $d$ channels increases, the gaps become more anisotropic, \cite{Hirschfeld:2011} and consequently $T_c$ decreases. An existing BS mode would then be damped strongly and hardly visible. This scenario could, in fact, apply for ${\rm Ba(Fe_{1-x}Co_x)_2As_2}$. If, on the other hand, the central hole bands disappear such as in ${\rm Rb_{0.8}Fe_{1.6}Se_2}$ or appropriately annealed FeSe\cite{HeSL:2013} the $d$ channel would prevail and nodeless $d_{x^2-y^2}$ pairing could be established. Since the gap is then quasi-isotropic $T_c$ can be comparably high as in the $s$ channel. From this point of view the transition temperatures in the cuprates are not yet maximal since the gap has nodes on the Fermi surface.
In any case, Raman scattering directly shows the symmetry of the competing pairing channels in ${\rm Ba_{0.6}K_{0.4}Fe_2As_2}$ and thus supports (i) the dominance of electronically driven pairing and (ii) shows directions in which higher transition temperatures may be expected.

\begin{acknowledgments}
We acknowledge useful discussions with Ming Yi. The work was supported by the DFG via the Priority Program SPP\,1458 (project no. HA\,2071/7) and, partially, via the Transregional Collaborative Research Center TRR\,80. Additional support came from the Bavarian Californian Technology Center BaCaTeC (project no. A5\,[2012-2]). Work in the Stanford Institute for Materials and Energy Sciences (SIMES) at Stanford and SLAC was supported by the US Department of Energy, Office of Basic Energy Sciences, Division of Materials Sciences and Engineering, under Contract No. DE-AC02-76SF00515.
\end{acknowledgments}


%

\newpage
\setcounter{figure}{0}
\makeatletter
\renewcommand{\thefigure}{A\@arabic\c@figure}
\makeatother

\begin{appendix}
\label{sec:appendix}

\section{Influence of the back-folding}
The simplest elementary cell has just one Fe atom per quadratic unit cell. This choice is motivated by the low-energy band structure of the FeSCs being derived only from Fe $3d$ orbitals. The resulting five bands reproduce the Fermi surfaces qualitatively but the magnetism cannot be treated appropriately. From the view point of light scattering the 1\,Fe cell proves sufficient for a qualitative understanding of the selection rules \cite{Muschler:2009}. However, the backfolding due to the
\begin{figure}[b]
  \centering
  \includegraphics[width=1.0\columnwidth]{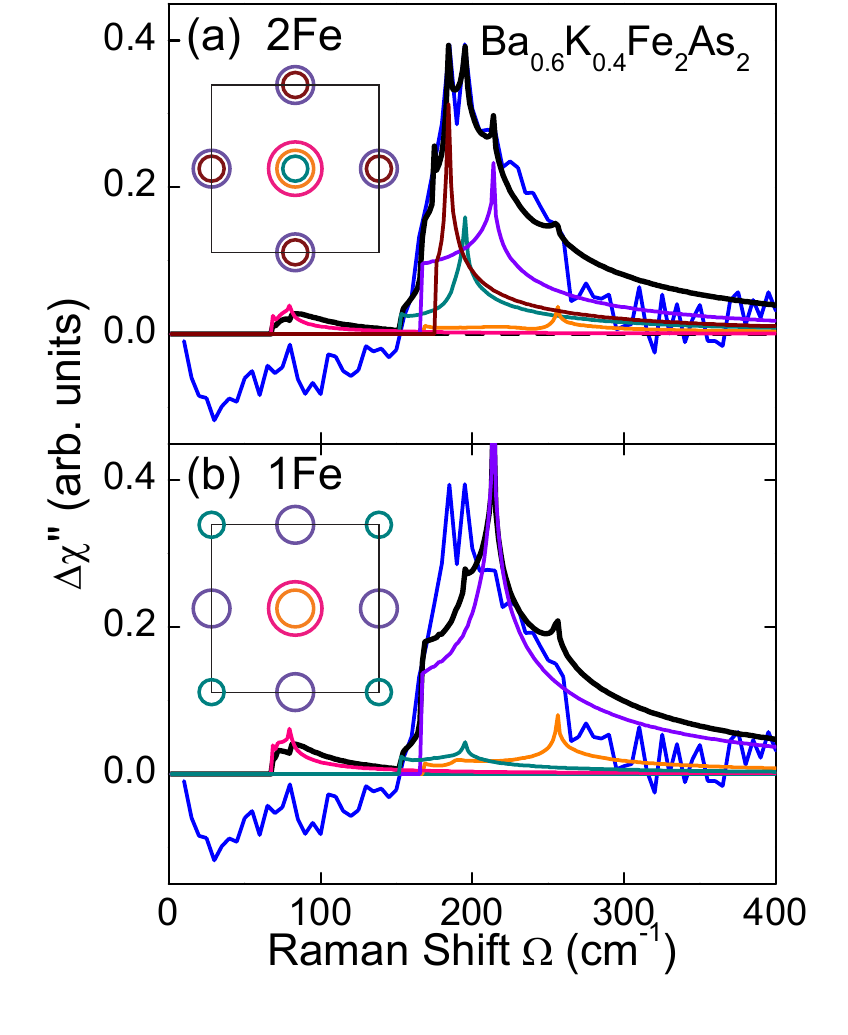}
  \caption{Results for $A_{1g}$ symmetry for (a) 1\,Fe and (b) 2\,Fe zone vertices. The gap parameters are unchanged. The insets show the bands for both cases. The difference between the $A_{1g}$ spectra is the biggest one of all symmetries.
  }
  \label{fig:1Fevs2Fe}
\end{figure} inclusion of the As atoms and the entire zone of $\rm BaFe_2As_2$, having a body centered tetragonal unit cell, changes the band structure considerably and influences also the selection rules \cite{Mazin:2010a}. In addition, the backfolding changes the spectral weight on the bands \cite{Ku:2011} further complicating the evaluation of one- and two-particle response functions. In our study we found good agreement upon using the band structure of the 2\,Fe unit cell. In addition to these calculations we redid some of the calculations in the 1\,Fe cell. In Fig.~\ref{fig:1Fevs2Fe} we show the results for $A_{1g}$ symmetry. While the overall shape is conserved there are minor but significant differences around the gap maximum. Therefore, if numerical studies are performed the 2\,Fe cell is preferable although the symmetry assignment is better done in the 1\,Fe cell since otherwise the generic meaning of the respective symmetries gets compromised. For instance, the $x^2-y^2$ symmetry being projected in the $B_{1g}$ spectra is the proper symmetry for both nematic fluctuations and the subdominant $d$ pairing channel discussed here. In the 2\,Fe cell one would have to switch to the $B_{2g}$ or $xy$ channel which appears awkward.

\begin{figure}[b]
  \centering
  \includegraphics[width=1.0\columnwidth]{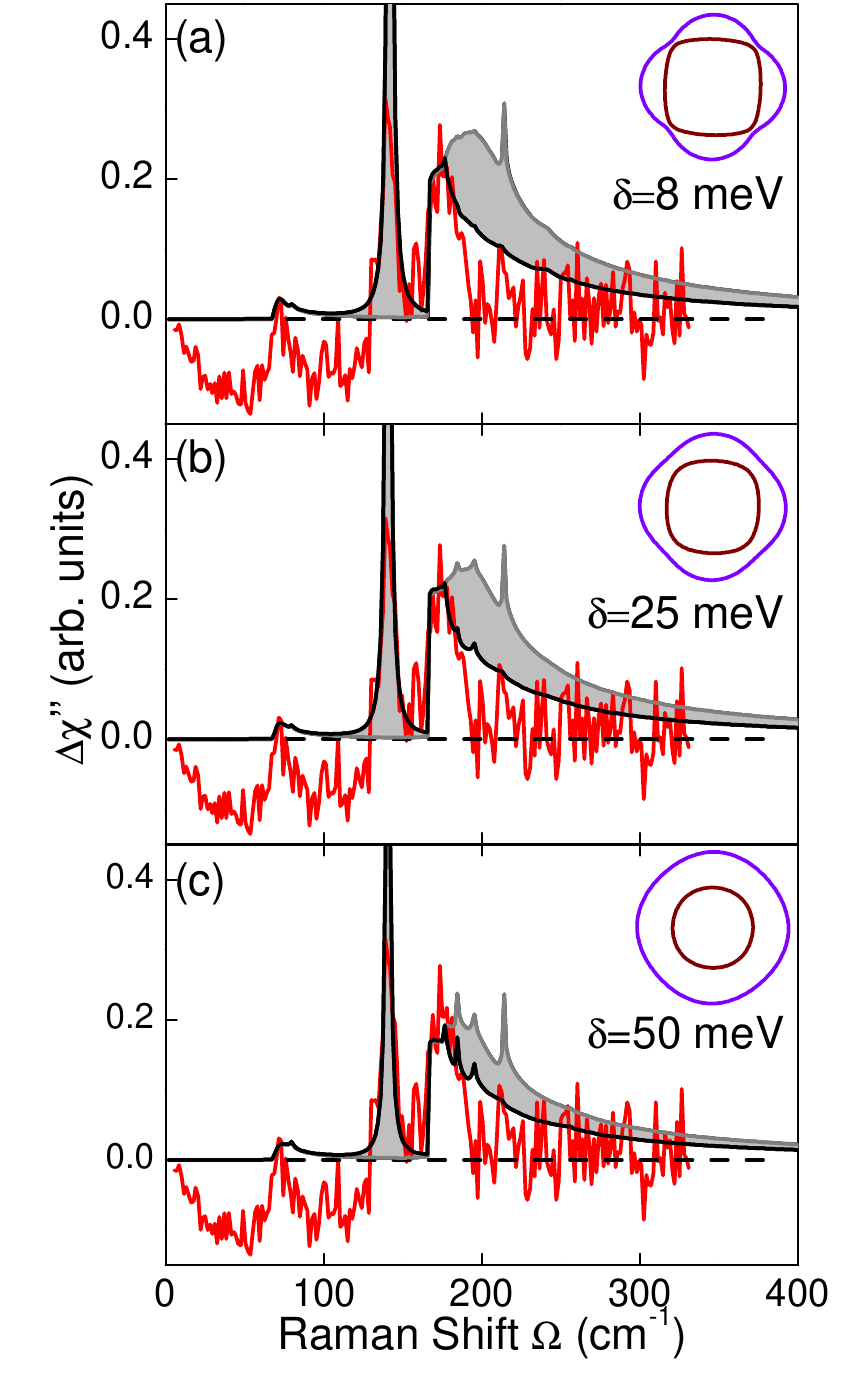}
  \caption{Results for $B_{1g}$ symmetry using different hybridization energies $\delta$ as indicated. The insets show a cut through the two electron bands at $k_{z} = \pi/2$.
  }
  \label{fig:hybridization}
\end{figure}
The backfolding makes the electron bands overlap. Since the electrons belong to the same orbitals the bands hybridize at the intersection points. As was shown by Mazin and coworkers \cite{Mazin:2010a} and by Eiter \textit{et al.} \cite{Eiter:2013} the cross section may be enhanced substantially at the hybridization point for the resulting increased band curvature and the spectra may change accordingly. Therefore, we also studied the effect of hybridization by calculating the $B_{1g}$ Raman spectra for various hybridization energies $\delta$ and plot the results in Fig.~\ref{fig:hybridization}. $B_{1g}$ is the most important symmetry in this context since the electron bands are the battle ground of the $s_{+-}$ and $d_{x^2-y^2}$ pairing channels. Although the Fermi surface shape clearly changes the spectra show only minor differences since the integrated spectral weight around the hybridization lines is almost independent of $\delta$ as opposed to the results for $\rm Ba(Fe_{1-x}Co_x)_2As_2$ \cite{Mazin:2010a}. We conclude that the influence of the hybridization does not complicate our argumentation. Rather the results are robust and show only small quantitative differences for the 1\,Fe and 2\,Fe basis.


\end{appendix}

\end{document}